\documentclass[12pt]{iopart}
\usepackage{graphicx}
\usepackage{amssymb}
\begin{document}
\title{Single microwave photon detection in the micromaser}
\author{M L Jones$^{1,2}$, G J Wilkes$^{2}$, and B T H Varcoe$^{1,2}$}
\address{$^1$ School of Physics and Astronomy, University of Leeds, Leeds, LS2 9JT, UK}
\address{$^2$ Department of Physics and Astronomy, University of Sussex, Brighton, BN1 9QH, UK}

\maketitle
\begin{abstract}
High efficiency single photon detection is an interesting problem for many areas of physics, including low temperature measurement, quantum information science and particle physics. 
For optical photons, there are many examples of devices capable of detecting single photons with high efficiency. However reliable single photon detection of microwaves is very difficult, principally due to their low energy. In this paper we present the theory of a cascade amplifier operating in the microwave regime that has an optimal quantum efficiency of 93\%. The device uses a microwave photon to trigger the stimulated emission of a sequence of atoms where the energy transition is readily detectable.  A detailed description of the detector's operation and some discussion of the potential limitations of the detector are presented.

\end{abstract}

The one atom maser or micromaser is an experiment where single atoms interact with a single mode of
the electromagnetic field of a resonant cavity \cite{Walther2006}, it therefore represents one of the most fundamental interactions
in quantum optics. 
The high Q cavity has a photon lifetime of up to half a second \cite{Walther2006} and the interaction between the atoms and the field generates a steady-state field in the 
cavity. 
The steady state cavity field in the micromaser has been the object of detailed experimental and theoretical studies for many years. 
Experiments include the observation of a sub-Poissonian statistical distribution of the field \cite{Rempe1990}, the quantum dynamics of the atom-field photon exchange represented in the collapse and revivals of the Rabi nutation \cite{rempe2}, atomic interference \cite{raithel1}, bistability and quantum jumps of the field \cite{Benson1994} and atom-field and atom-atom entanglement \cite{englert}. 
The micromaser has also been used to to observe coherent population trapping of photon Fock states \cite{weidinger99} and Fock state Rabi oscillations \cite{varcoe00}.
Photon trapping states are a direct result of field quantisation and represent the extreme state of a field as a boson. 
These states have proven to be so experimentally robust that they can now be used as a tool to achieve other goals. 
One application was the creation of a single atom/single photon source \cite{Brattke2001}, an essential item in an atomic beam quantum computer \cite{Blythe2006}. 
In this paper we report on another application of coherent population trapping of Fock states, where they can be used in the operation of a high efficiency photon detector. 
 
A single photon detector in the microwave regime is particularly interesting because single microwave photons are very difficult to detect, principally due to the extremely low energy of a photon (around $10^{-5}$eV). 
However, using the micromaser it is possible to achieve a cascade amplification of microwave photons which can provide an unambiguous, time resolved detection event. 
Strong coupling in cavity QED has a very strong single photon interaction strength and therefore several mechanisms exist for detecting single microwave photons in cavity QED. One example is the non-demolition probe used in ref.\ \cite{Gleyzes2007} to detect single microwave photons in a microwave cavity. The detector presented in this paper gives a similar time resolution to ref.~\cite{Gleyzes2007}, with possible advances in atomic pumping schemes giving further improvements.

\begin{figure}[ht] 
\begin{center}
\includegraphics[width=0.48\textwidth]{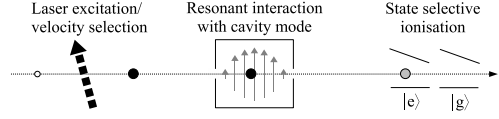} 
\caption{Schematic of operation of the micromaser. Ground state atoms (unfilled circle) exit the oven with thermal velocities. A
detuned angled laser excitation region excites a particular velocity
class to the Rydberg state $\vert e\rangle$ (black circles). The transition
between this and a second Rydberg state $\vert g\rangle$ is resonant with a single mode in
a superconducting microwave cavity and interacts coherently with it. The
atomic state is recorded by state selective field ionization detectors 
upon exiting the cavity.}
\label{fig:micromaser} 
\end{center} 
\end{figure}

The micromaser (Fig.~\ref{fig:micromaser}) uses rubidium-85 atoms laser excited
to the Rydberg $63P_{3/2}$ state, which acts
as the upper level (excited state) $\vert e\rangle$ a two level
system. The lower level (ground state) $\vert g\rangle$ is the Rydberg $61D_{5/2}$ state with an $\vert e\rangle\leftrightarrow\vert g\rangle$ transition frequency of 21.456 GHz. 
However Rydberg--Rydberg transitions over a range of frequencies from 10--120GHz are readily accessible.
The excitation laser is angled with respect to the atomic beam to allow velocity selection via Doppler detuning. Using this technique, combined with a time-of-flight scheme \cite{Hagley1997}, a typical velocity resolution of 0.5\% is achievable. 
The excited atoms enter the high-$Q$ superconducting cavity and interact
resonantly with a TE$_{121}$ mode of the
resonator via the Jaynes-Cummings Hamiltonian, 

\begin{equation}\label{eq:jaynescummings}
\hat{H}=\frac{1}{2}\hbar\omega_{0}\hat{\sigma}_{z}+\hbar\omega\hat{a}^{
\dagger}\hat{a}+\hbar
g\left(\hat{a}^{\dagger}\hat{\sigma}_{-}
+\hat{a}\hat{\sigma}_{+}\right),
\end{equation}
where $\omega_{0}$ and $\omega$ are the atomic transition and field mode
frequencies respectively, $\hat{\sigma}_{z}$ is the atomic projection
operator, $\hat{\sigma}_{\pm}$ are the atomic raising and lowering
operators, $g$ is the coupling strength ($\approx 40\mathrm{krads^{-1}}$
in the micromaser) between atom and field, $\hat{a}$ and
$\hat{a}^{\dagger}$ are the photon annihilation and creation operators.
This is one of the simplest Hamiltonians in quantum optics, describing
the interaction of a single two-level atom with a single field mode.
There are no loss mechanisms in this ideal model, but since the
interaction time is 3--4 orders of magnitude smaller than the time
scales for losses (the micromaser cavity has a $Q$-factor of up to
$5\times 10^{10}$), this is an excellent approximation.

During the interaction, the atom undergoes Rabi oscillations between
the states $\vert e\rangle$ and $\vert g\rangle$. After exiting the
cavity, the state of the atom is measured by state selective field
ionization, giving us information about the field. The probability that
an atom emits a photon into the cavity is given by 
\begin{equation}
P_{\mathrm{emit}}=\sin^{2}\left(g \tau\sqrt{n+1}\right),\label{eq:prob}
\end{equation} 
where $\tau$ is the interaction time and $n$ is the
number of photons already in the cavity. By tuning the atomic velocity
correctly, we can reduce this probability to zero, which can be
understood as the system undergoing an integer number of Rabi
oscillations, 
\begin{equation}
\tau=\frac{k\pi}{g\sqrt{n+1}},\label{eq:trapping} 
\end{equation} 
where $k$ is an integer. If equation \ref{eq:trapping} is fulfilled, then the cavity field with $n$ photons has zero probability of progressing to $n+1$. These are the trapped photon states \cite{weidinger99, varcoe00} and they can be accessed by tuning the interaction time. 
Under these conditions, the emission probability can change dramatically. 
For example when the interaction time is chosen so that the micromaser is operating in the vacuum ($n=0$) trapping state, the emission probability changes from zero to 93\% on the addition of a single photon. 
One photon therefore has a dramatic and detectable effect on the atomic statistics producing a significant change in ground state count rate with a cascade of $\vert g\rangle$ detector counts. 

\begin{figure}[ht] 
\begin{center}
\includegraphics[width=0.48\textwidth]{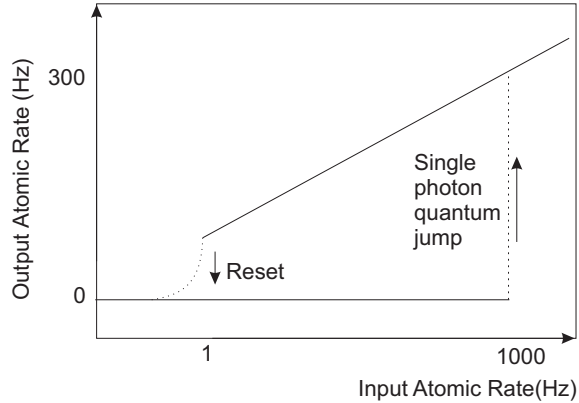} 
\caption{The basic principle of the detector is operating the micromaser in a bistable mode with a strong hysteresis curve. The transition from the lower arm to the upper arm is triggered by a single photon. Reducing the pump strength returns the cavity to the lower branch of the hysteresis curve. The input atomic rate is determined by the excitation laser frequency and intensity and is controllable. The output atomic rate is determined by the cavity decay rate and the micromaser dynamics. }
\label{fig:hyst} 
\end{center} 
\end{figure}

Fig.~\ref{fig:hyst} demonstrates the basic principle of the detector. 
In the coherent trapping regime there is a strong hysteresis between two macroscopically different bistable states and both states are simultaneously solutions to the operating conditions.
The lower branch is unstable on the addition of a photon causing a spontaneous jump to the upper branch. This jump is detectable by looking at the flux of ground state atoms.
In the lower arm the rate of ground state atoms emerging from the cavity is identically zero and the micromaser remains in this state until the arrival of a photon. Upon the arrival of the photon there is a discontinuous change in atomic emission probability from zero to 93\% and the ground state atom rate jumps to the upper branch. 
The cavity is reset either by reducing the pump atom rate and allowing the cavity to return to the lower arm of the hysteresis curve or with a pulse of ground state atoms that can absorb the photon field.

Quantum Trajectory Analysis \cite{cresser96} produces experimentally realistic detection records and allows us to perform a simulation of the micromaser using realistic operating conditions.
It involves stochastically evolving a wavefunction using a combination of a
non-Hermitian Hamiltonian and a set of ``jump operators''. 
We use this method to perform individual trajectories and hence to create theoretical results that
simulate real detection records from the experiment.
The non-Hermitian effective Hamiltonian is as in \cite{cresser96}, given by
\begin{equation}
\hat{H}_{\mathrm{eff}}=-\frac{1}{2}i\hbar\gamma
\left[\left(n_{t}+1\right)\hat{a}^{\dagger}\hat{a}+n_{t}\hat{a}\hat{a}^{
\dagger}\right]-\frac{1}{2}i\hbar
R+\hbar\omega\hat{a}^{\dagger}\hat{a},
\end{equation}
where $\gamma$ is the cavity decay constant, $n_{t}$ is the
thermal photon number and $R$ is the rate at which atoms pass through
the cavity.

The particular set of jump operators used are as in \cite{cresser96}, listed below:
\begin{equation}
\hat{C}_{-1}=\sqrt{\gamma \left(n_{t}+1\right)}\hat{a}
\end{equation}
is the operator that represents a photon being lost to the
reservoir,
\begin{equation}
\hat{C}_{0}=\sqrt{R}\cos\left(g\tau\sqrt{n+1}\right)
\end{equation}
represents an atom exiting the cavity (containing $n$ photons) in the excited state and leaving the photon number unchanged,
\begin{equation}
\hat{C}_{1}=\sqrt{R}\frac{\sin\left(g\tau\sqrt{n}\right)}{\sqrt{n}}
\hat{a}^{\dagger}
\end{equation}
is the operator representing an atom introducing a photon into the field
that contains $n$ photons and exiting in the ground state, and
\begin{equation}
\hat{C}_{2}=\sqrt{\gamma n_{t}}\hat{a}^{\dagger}
\end{equation}
is the operator representing a photon being gained from the reservoir.
However, since every operator maps pure states onto pure states, and for
our purposes we always begin with the (pure) vacuum state, then we can
reduce the dynamics simply to jumps occurring stochastically and the
wavefunction remaining unchanged in between.

\begin{figure}[ht]
\begin{center}
\includegraphics[width=0.48\textwidth]{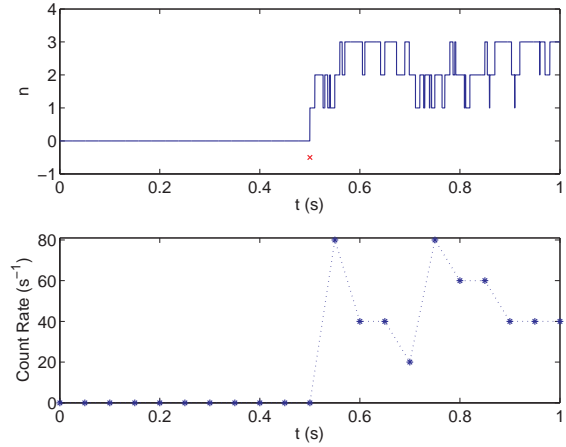}
\caption{A pair of graphs showing how the change in the cavity
photon number affects the ground
state count rate for ideal detectors. In the upper plot, the cross at $t=0.5s$ denotes the time that a photon arrives in the cavity and its subsequent effect upon the photon number. The lower plot shows how this affects the observed count rate. Notice that a very small
change in the cavity photon number can produce a very large change in
count rate. In this example, the atomic flux was set to $R=100$ and the cavity decay rate $\gamma=20$.}\label{fig:results1}
\end{center}
\end{figure}

A quantum trajectory simulation of the micromaser operating with $\tau=\tau_0=\pi/g$ was performed, in which the ground state detector
count (the rate of occurence of jump $\hat{C}_2$ measured weighted by detector efficiency $\eta_g$) was monitored while
single photons were added to the cavity at random times. Figure
\ref{fig:results1} shows an example trajectory for the ideal case, with
no deviation from perfect operating conditions, in order to illustrate
the principle of the operation of the detector. It shows how the field
evolves inside the cavity, along with the detector clicks we see when
probing the atoms.
For sufficiently high atomic pump rates (in this case, $R=100$s$^{-1}$), once the
$n=0$ state is passed then the field very quickly reaches
three photons, which also gives a zero emission probability and emission is forbidden once again (this is the ($n=3$, $k=2$) state,
eq.~\ref{eq:trapping}). The field then proceeds to rapidly oscillate
between this and the two and one photon states giving rise to the high count rate. 
In other words, adding just one
photon produces a change in detector count rate of up to 80 counts per second, which is easily detectable, even in the presence of noise. 
\begin{figure}[ht]
\begin{center}
\includegraphics[width=0.48\textwidth]{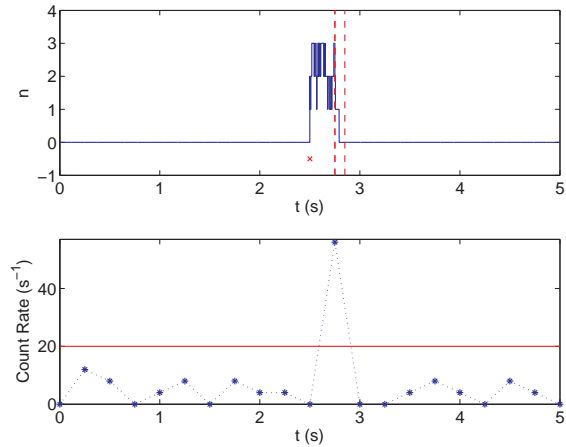}
\caption{This figure shows the normal operation of the detector. When the count rate exceeds the predefined threshold, a detection event 
is recorded and the cavity field is allowed to relax back to the vacuum
state by switching off the excitation laser. The pair of dashed vertical lines on the top plot shows the period during which the laser excitation was switched off to allow the field to decay back to the vacuum state. This could be performed more quickly by using a pulse of ground state atoms to clean the cavity mode.}
\label{fig:threshold}
\end{center}
\end{figure}
Figure \ref{fig:threshold} displays the threshold operation with a background count rate of 4s$^{-1}$ and threshold of 20s$^{-1}$. 
Once a detection has occurred, the field is then allowed to relax back to the vacuum state, either via free decay
of the field, delineated by vertical dashed lines in figure \ref{fig:threshold} or using a ground state atom mode cleaning pulse.

While this model is created for ideal conditions, it is also possible to include departures from the ideal conditions in the quantum trajectory method to
investigate the limits imposed on the system. 
The factors include; dark counts, caused by detector
clicks occurring without an associated Rydberg ionization event, leading to a small background count rate of less than 3 Hz; missed counts, caused by less than 100\% detection efficiency and misidentification
when an atom is ionized at the wrong detector.
These errors are incorporated into the model via
the detector efficiencies $\eta_g<1$ and $\eta_e<1$ for the ground and
excited state detectors respectively and adding a random background signal
generated with a poissonian distribution centred at $r_{b}$
counts/second to simulate the dark counts and crosstalk.

\begin{figure}
\begin{center}
\includegraphics[width=0.48\textwidth]{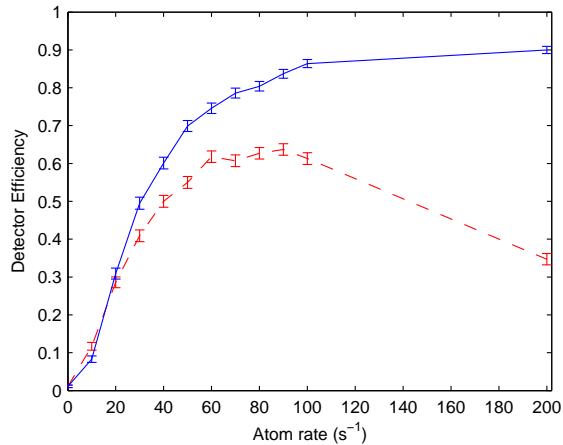}
\caption{Plot of detector efficiency for different values of $R$, with
associated error bars. The broken line shows the efficiency for high numbers of two atom events and the solid line shows the results for normal operation.
The parameter values are: $r_{b}=2$, $\Delta\phi/\phi=0.5\%$, $\eta_g=0.8$, 
$\mathrm{threshold} = 10$.}\label{fig:efficiency}
\end{center}
\end{figure}

Other errors in the system arise from the departure from ideal operating
conditions of the micromaser. The idealised model assumes that there is
no spread in interaction time $\tau$, the coupling parameter $g$ is
constant and that there is never more than one atom in the cavity at any
time. 
In practice, we find that, due to the linewidths of the velocity
selecting laser and atomic transition, the interaction time has a
non-zero spread. 
Mechanical vibrations in the system may also cause
variations in the parameter $g$. 
Hence we replace $g$ and $\tau$ with
$\phi=g\tau$, drawn from a normal distribution centred at $\phi_{0}=\pi$
with spread $\Delta\phi$ to represent these effects. 
Finally, as the pump atoms are thermally distributed, there is a non-zero probability of a two atom event occurring given by $P=e^{-R\tau}$.
This effect is discussed elsewhere in the literature \cite{Kist1996,Ariunbold2000} and can also be included in the simulation and the resulting model is presented in figure \ref{fig:efficiency} which compiles the total efficiency of dectection that we can expect.

In this paper we have shown that it is possible to design a high quantum efficiency single microwave photon detector.
Figure \ref{fig:efficiency} shows the efficiencies that are achieved for a range of atomic rates $R$ for atomic beams with low numbers of two atom events (solid line) and for conditions in which we have high rates of two atom events (broken line). The detection efficiency under normal operation reaches around $93\%$ at values of $R>500$. 
The decrease in efficiency observed in the high two atom event case is due to the destruction of the Jaynes Cummings dynamics when two atoms are present in the cavity at the same time. This demonstrates that limiting two atom events will therefore be a very important criteria in the experimental design. This can be achieved with careful design of the cavity and balancing the atomic pump rate and cavity decay time.   
It is possible for the detector to be designed with a small deadtime and good timing resolution ($<<1$s). 
For events with a low average photon number, such as in the detection of black body thermal photons, this would be sufficient to determine the cavity temperature to high accuracy and would therefore have applications in precision temperature measurement. The problem of coupling external photons into the cavity remains an open problem, but a promising approach has been shown in the task of single photon transfer in a quantum network \cite{Cirac1998}, whereby time symmetric pulse shaping can optimise the coupling efficiency.
 
\ack 
This work was funded by the White Rose Universities Consortium and Leeds University.

\end{document}